\documentclass[useAMS, usenatbib, usegraphicx]{mn2e}

\newcommand{\kw}{Konus-\textit{Wind}\ }

\title[]
  {A search for giant flares from soft gamma-repeaters in nearby galaxies in 
the \kw short burst sample}
\author[Svinkin et al.]
  {D.~S.~Svinkin,$^1$ K.~Hurley,$^2$ R.~L.~Aptekar,$^1$ S.~V.~Golenetskii,$^1$ 
D.~D.~Frederiks$^1$ \\
$^1$Ioffe Physical-Technical Institute, Politekhnicheskaya 26, 
St.~Petersburg, 194021, Russia\\
$^2$Space Sciences Laboratory, University of California, 
7 Gauss Way, Berkeley, CA 94720-7450, USA
}

\pagerange{\pageref{firstpage}--\pageref{lastpage}} \pubyear{2014}

\begin{document}

\label{firstpage}

\maketitle

\begin{abstract}
The knowledge of the rate of soft gamma-ray repeater~(SGR) giant flares is important for 
understanding the giant flare mechanism and the SGR energy budget in the 
framework of the magnetar model. We estimate the upper limit to the rate using the 
results of an extensive search for extragalactic soft gamma-repeater giant flares (GFs) 
among 140 short gamma-ray bursts detected between 1994 and 2010 by \kw using 
InterPlanetary Network (IPN) localizations and temporal parameters. 
We show that \kw and the IPN are capable of detecting GFs with energies of $2.3\times10^{46}$~erg 
(which is the energy of the GF from SGR~1806$-$20 assuming a distance of 15~kpc) 
at distances of up to $\sim 30$~Mpc and GFs with energies of $\la 10^{45}$~erg 
(which is the energy of the GF from SGR~0526$-$66) at distances of up to $\approx 6$~Mpc. 
Using a sample of 1896 nearby galaxies we found that only two bursts, GRB~051103 and 
GRB~070201, have a low chance coincidence probability between an IPN localization and 
a nearby galaxy. We found the upper limit to the fraction of GFs among short GRBs 
with fluence above $\sim 5\times 10^{-7}$~erg~cm$^{-2}$ to be $< 8$\% 
(95\% confidence level). Assuming that the number of active SGRs 
in nearby galaxies is proportional to their core-collapse supernova rate, we derived 
the one-sided 95\% upper limit to the rate of GFs with energy output similar to
the GF from SGR~1806$-$20 to be $(0.6\textrm{--}1.2)\times 10^{-4} Q_{46}^{-1.5}$~yr$^{-1}$~per~SGR, 
where $Q_{46}$ is the GF energy output in $10^{46}$~erg. 
\end{abstract}


\section{INTRODUCTION}
Soft gamma repeaters (SGRs) are thought to be a special rare class of neutron stars 
exhibiting two types of bursting emission in gamma-rays. During the active stage 
SGRs emit randomly occurring short ($\sim 0.01$--1~s) hard X-ray bursts with peak 
luminosities of $\sim 10^{38}\textrm{--}10^{42}$~erg~s$^{-1}$. The bursting activity 
may last from several days to a year or more, followed by a long quiescent period. 
Much more rarely, perhaps once during the SGR stage, an SGR may emit a giant flare (GF) 
with sudden release of an enormous amount of energy $\sim(0.01\textrm{--}1)\times10^{46}$~ergs. 
A GF displays a short $\sim 0.2$--0.5~s initial pulse of gamma-rays with a sharp rise 
and a more extended decay which evolves into a soft, long-duration decaying tail modulated 
with the neutron star rotation period; see~\citet{Mereghetti2013} for a detailed review. 


To date, 15 SGRs have been discovered~\citep{Olausen_Kaspi2014}, 14 of which are in our Galaxy and one is 
located in the Large Magellanic Cloud (LMC). The first GF was detected from the 
LMC on March~5, 1979 by the Konus experiment on the Venera~11 and~12 interplanetary 
missions~\citep{Golenetskii1979SvAL, Mazets1979} as well as by the 
Interplanetary Network (IPN)~\citep{Barat1979, Cline1980, Evans1980, Cline1982}. 
So far, only three GFs have been observed and all of them were detected by the 
Konus experiments and the IPN. Two recent GFs, from SGR~1900+14 and SGR~1806$-$20, were 
preceded by a pronounced increase in bursting activity~\citep{Mazets1999a, Frederiks2007}.

All of the known SGRs are rapidly spinning down X-ray pulsars with spin periods 
of 2--12~s and persistent X-ray luminosities of $10^{35}$--$10^{36}$~erg~s$^{-1}$. 
The SGRs are thought to belong to a wider class of objects called magnetars. 
This class also includes anomalous X-ray pulsars (AXPs) and high-B radio pulsars; 
a recent catalogue lists 26 objects in the magnetar class~\citep{Olausen_Kaspi2014}. 
The distinction between AXPs and SGRs is blurred due to the detection of anomalously 
high quiescent X-ray luminosity of SGRs and the occasional bursting of AXPs. 
About half of the known magnetars are associated with star forming regions or SNRs 
and appear to be young isolated NS. The spatial distribution of magnetars provides 
evidence that they are born from the most massive O stars~\citep{Olausen_Kaspi2014}. 
It is believed that the activity of magnetars is related to the presence of 
a superstrong magnetic field $\sim10^{13}$--$10^{15}$~G, inferred from the high 
spin down rate and other evidence~\citep{Duncan_and_Thompson_1992ApJ, Thompson_and_Duncan_1995MNRAS, 
Thompson_and_Duncan_1996ApJ}. The first direct measurement of the high spin down rate 
of SGR~1806$-$20 by~\citet{Kouveliotou1998} gave strong support to the magnetar nature of SGRs.

Due to the enormous luminosity of the initial pulse GFs can be detected from SGRs 
in nearby galaxies. In this case, the initial pulse will be indistinguishable from a short 
gamma-ray burst (GRB). The rate of soft gamma-ray repeater GFs and their fraction 
among short-duration GRBs are important quantities for the understanding of the giant flare mechanism 
in the framework of the magnetar model. A limit on the fraction 
of SGR GFs among short GRBs was estimated in several 
studies~\citep{Lazzati2005, Nakar2006ApJ, Palmer2005, Popov2006, Ofek2007} 
to be $\sim 1$--15\%; see~\citet{Hurley2011} for a detailed review. Two extragalactic GF 
candidates have been reported and localized by the IPN:
GRB~051103 in the the M81/M82 group of galaxies~\citep{Ofek2006, Frederiks2007a, Hurley2010} 
and GRB~070201 in the Andromeda galaxy~\citep{Mazets2008, Ofek2008}. However, \citet{Hurley2010} 
argued that GRB~051103 is unlikely to be an SGR giant flare in the nearby Universe, 
mostly due to its extreme peak luminosity of approximately $4.7\times10^{48}$~erg~s$^{-1}$, 
assuming it was from an SGR in the M81/M82 group, which is a factor of 10 brighter 
than the peak luminosity of the giant flare from SGR~1806$-$20, which is 
$(2\textrm{--}5)\times 10^{47}$~erg~s$^{-1}$ assuming a distance of 15~kpc.

A recently published catalogue of IPN localizations of \kw short GRBs~\citep{Palshin2013} 
contains localizations of 271 short GRBs detected by \kw during its almost 16~yrs of continuous 
full-sky observations. This catalogue allows us to search for GF candidates in the 
largest well-localized short GRB sample. 

In Section~\ref{KW_sensitivity} we discuss the \kw and IPN sensitivity to GFs. In Section~\ref{Gal_sample} 
we provide a nearby galaxy sample and discuss its properties. We present in Section~\ref{GF_search} 
the search for \kw short GRBs that spatially coincide with the nearby galaxies and 
estimate the fraction of GFs among short GRBs. In Section~\ref{GF_rate} we use 
results of this search to derive upper limits on the GF rate. 
Finally, in Section~\ref{Discussion} we give concluding remarks.

\section{THE KONUS-\textit{WIND} AND IPN SENSITIVITY TO GIANT FLARES} \label{KW_sensitivity}
The \kw instrument~\citep{Aptekar1995SSR} consists of two identical omnidirectional NaI(Tl) 
detectors. The detectors are mounted on opposite 
faces of the rotationally stabilized \textit{Wind} spacecraft. Each detector has 
an effective area of $\sim 80$--160~cm$^{2}$ depending on the photon energy and incidence angle. 
\kw has two operational modes: background and triggered. The triggered mode is 
initiated when the number of counts on one of two fixed time scales, 1~s or 140~ms, 
exceeds a preset value. This value is defined in terms of a number of standard 
deviations above the background ($\approx 9\sigma$). There is a single background 
measurement period of 30~s duration before the trigger interval. The initial 
pulse of a GF from a distant galaxy can trigger \kw on the 140~ms time scale.

Among the three detected GFs, only one, from SGR~1806$-$20, has well estimated 
initial pulse spectral parameters. The parameters were derived using the detection 
of the flare radiation reflected from the Moon~\citep{Frederiks2007}. The initial 
pulse spectrum of this GF is well described by a power law with an exponential 
cutoff ($dN/dE \sim E^{-\alpha}\exp(-E/E_0)$) with ${\alpha=0.73^{+0.47}_{-0.64}}$ 
and~${E_{0}=666^{+1859}_{-368}}$~keV (or the $\nu F_{\nu}$ spectrum peak energy 
${E_{\rmn{p}} = (2-\alpha)E_0 = 850^{+1259}_{-303}}$~keV), see also~\citet{Terasawa2005}. 
Two other GFs are characterized by only rough estimates 
of $E_{\rmn{p}}$: ${\sim 400\textrm{--}500}$~keV (SGR~0526$-$66,~\citealp{Golenetskii1979SvAL, Mazets1979}) 
and~$>250$~keV (SGR~1900+14,~\citealp{Hurley1999, Mazets1999}).

Assuming all the initial pulse energy is released during the short trigger 
timescale (140~ms), we calculated the minimum fluence $S_{\rmn{min}}$ in the 20--10000~keV 
band which produces a $9\sigma$ increase in the count rate in the 50--200~keV energy 
range using forward folding of an incident photon spectrum through the detector response matrix. 
For calculations we used an average background count rate of 400~counts~s$^{-1}$ 
in the 50--200~keV energy range. 

\begin{figure}
\includegraphics[width=84mm]{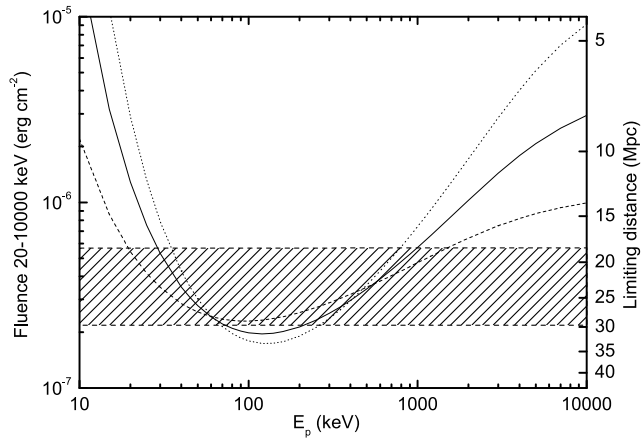}
\caption{Dependence of the 140~ms fluence (20~keV--10~MeV) and the limiting 
distance ($d_{\rmn{max}}$), assuming $Q=2.3\times 10^{46}$~erg, on GF spectral parameters $E_{\rmn{p}}$ 
and $\alpha$: $\alpha=1$ (solid line), $\alpha=1.5$ (dashed line), $\alpha=0.5$ (dotted line). 
The limiting distance for bursts with $E_{\rmn{p}} \sim 250\textrm{--}1000$~keV is in the range  
$\sim 18$--30~Mpc (dashed region).}
\label{fig1}
\end{figure}

We found that $S_{\rmn{min}}$ strongly depends on the burst spectral hardness ($E_{\rmn{p}}$ and $\alpha$).
The known $E_{\rmn{p}}$ range of about 200--1000~keV  corresponds to 
${S_{\rmn{min}}= (2.1\textrm{--}5.7)\times 10^{-7}}$~erg~cm$^{-2}$~(Fig.~\ref{fig1}). 
Then, the corresponding limiting detection distance was calculated as 
${d_{\rmn{max}} = \sqrt{Q/(4\pi S_{\rmn{min}})}}$, where $Q$ is the GF energy output. 

The IPN~\citep{Hurley2013EAS} is a set of gamma-ray detectors on 
board several spacecraft orbiting the Earth, Mars and Mercury. The IPN allows 
near continuous, all-sky coverage and provides accurate sky locations determined by 
triangulation between the spacecraft in the network using GRB photon arrival times. 
Each pair of spacecraft constrains the GRB position to an annulus on 
the celestial sphere, and the intersection of two or more annuli forms an error box. 
Depending upon the spacecraft involved, the error box areas can vary 
from square arcminutes to thousands of degrees.

\begin{figure}
\includegraphics[width=84mm]{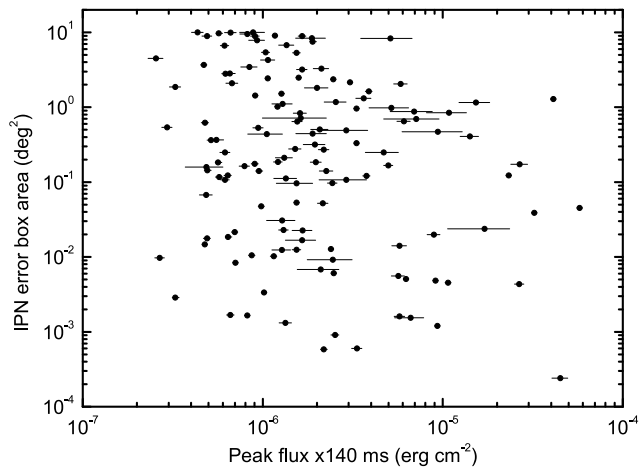}
\caption{IPN error box area versus \kw fluence (20--10000~keV) over the 140~ms 
interval with the highest count rate.}
\label{fig2}
\end{figure}

We investigated the dependence of the IPN error box area on the \kw fluence (20--10000~keV) 
over the 140~ms interval with the highest count rate and found no prominent 
correlation (Fig.~\ref{fig2}). So the IPN sensitivity to short GRBs 
in terms of the error box area does not depend on the \kw burst fluence. 
Therefore $S_{\rmn{min}}$ can be taken as the IPN sensitivity to short GRBs.

The distance estimates for SGR~1806$-$20 range from about 6~kpc to 19~kpc~\citep{Tendulkar2012ApJ};
the most recent analysis of~\citet{Svirski2011} gives the range 9.4--18.6~kpc. 
The isotropic energy release of the initial pulse of the GF from SGR~1806$-$20 
(${Q = 2.3\times 10^{46} d_{15}^2}$~erg) implies a limiting distance 
${d_{\rmn{max}} = (18\textrm{--}30)\times d_{15}}$~Mpc, 
where~$d_{15}=d/15\textrm{~kpc}$ and~$d$ is the distance to SGR~1806$-$20. 
The range of~$d_{\rmn{max}}$ corresponds to the~$S_{\rmn{min}}$ range.
Less intense GFs with ${Q\approx 1 \times 10^{45}}$~erg (like the 1979 March~5 GF from SGR~0526$-$66) 
can be detected up to~${d_{\rmn{max}} = (3.8\textrm{--}6.3) Q_{45}^{0.5}}$~Mpc, 
where $Q_{45}$ is the GF energy output in $10^{45}$~erg. 

\section{THE GALAXY SAMPLE} \label{Gal_sample}
The Gravitational Wave Galaxy Catalogue~(GWGC, \citealp{White2011CQGra}) contains 
over 53,000 galaxies within 100~Mpc. Our initial sample includes 8112 of them 
within a distance 30~Mpc. The distance uncertainty in the sample ranges from 15\% to~22\%. 
Using the method suggested by~\citet{Ofek2007} we estimate the completeness of 
the galaxy sample due to obscuration by the Galactic plane to be $\epsilon_{\rmn{G}}=94$\%. 
Under the assumption that all SGRs are young isolated neutron stars we postulate 
that the number of SGRs is proportional to the galaxy core-collapse supernova 
(CCSN, SN types Ib/c and II) rate.

Following~\citet{Cappellaro1999} and~\citet{Boser2013} we assume that 
the galaxy's blue luminosity $\rmn{L}_{B}$ 
is a good estimator of the galaxy CCSN rate $R_{\rmn{SN}} = k L_{B}$, where 
$k$ is a factor depending on the galaxy type: 
${k_{\rmn{E-S0}}<0.05}$, 
${k_{\rmn{S0a-Sb}}=0.89\pm0.33}$, 
${k_{\rmn{Sbc-Sd}}=1.68\pm0.60}$, 
${k_{\rmn{Sm, Irr., Pec.}}=1.46\pm0.71}$, 
where $k$ is given in SNu ($1\rmn{SNu}={1\rmn{SN}(100\rmn{yr})^{-1}(10^{10} \rmn{L}_{\sun B})^{-1}}$) and 
$\rmn{L}_{\sun B} = 2.16\times10^{33}$~erg~s$^{-1}$ 
is the Sun's blue luminosity. The $\rmn{L}_{B}$ of a galaxy was calculated as 
$\rmn{L}_{B}=10^{-0.4(\rmn{M}_B-\rmn{M}_{\sun B})} \rmn{L}_{\sun B}$, 
where $\rmn{M}_{\sun B}=5.48$ is the Sun's absolute blue magnitude. 

The initial galaxy sample contains 790 galaxies without any given $\rmn{L}_{B}$, 
so the $\rmn{L}_{B}$ sample completeness is about $\epsilon_{\rmn{L}}=90$\%. Among the 7322 
galaxies with a given $\rmn{L}_{B}$, 2405 have an unknown morphological type. 
These galaxies are, on average, about three magnitudes dimmer than those with a given 
type and they contain less than 7\% of the cumulative supernova rate, 
so we do not take them into account. Then we limited our final sample to 1896 galaxies with 
the highest $R_{\rmn{SN}}$ which contain $\epsilon_{\rmn{SN}}=90$\% of the cumulative $R_{\rmn{SN}}$ 
and have morphological types other than E and S0. The sky surface density of these galaxies 
is 0.046~deg$^{-2}$. The cumulative supernova rate for this sample 
is~$R_{\rmn{SN}}={22.8\pm0.4}$~yr$^{-1}$.

\begin{figure}
\includegraphics[width=84mm]{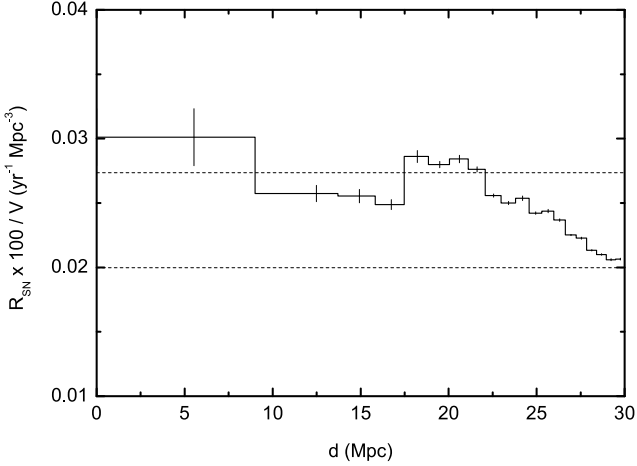}
\caption{Distance dependence of the local volumetric CCSN rate $R_{\rmn{SN}}/V$, 
where $V$ is a volume within distance $d$. 
The solid line is $R_{\rmn{SN}}/V$ for 1896 galaxies within 30~Mpc. 
The horizontal dashed lines denote $\pm 1\sigma$  interval for the local CCSN rate from~\citet{Mattila2012},
 $1.9_{-0.2}^{+0.4} \times 10^{-4}$~yr$^{-1}$~Mpc$^{-3}$ assuming a 
 missing CCSN fraction of 0.189.}
 \label{fig3}
\end{figure}

We compared the estimated volumetric $R_{\rmn{SN}}$ for the 1896 galaxies with the lower 
limit for the local CCSN rate (distances up to 15~Mpc) derived by~\citet{Mattila2012} 
which is ${1.9_{-0.2}^{+0.4}\times 10^{-4}}$~yr$^{-1}$~Mpc$^{-3}$. 
The volumetric $R_{\rmn{SN}}$ is shown in Fig.~\ref{fig3}. The volumetric $R_{\rmn{SN}}$ derived from 
the galaxy blue luminosity is consistent within $1\sigma$ with the observed CCSN rate 
assuming that $\approx 19$\% of local CCSN were missed by the optical surveys. 
The volumetric CCSN rate shows a decline beyond $\sim 22$~Mpc. The decline 
could be the result of observational selection effects, therefore we used the
volumetric CCSN rate averaged over the distance range up to 22~Mpc, 
which is ${(2.74\pm0.18) \times 10^{-4}}$~yr$^{-1}$~Mpc$^{-3}$ ($1\sigma$ CL), 
as an estimate of the CCSN rate at larger distances.

We also found an excess in the volumetric CCSN rate within $\sim 10$~Mpc up to 
${(9.3\pm1.6) \times 10^{-4}}$~yr$^{-1}$~Mpc$^{-3}$ ($1\sigma$ CL) within 5.1~Mpc, 
with only five galaxies containing 25\% of the total CCSN rate: 
PGC047885 at $d = 5$~Mpc, IC~0342 at 3.28~Mpc, NGC~6946 (Fireworks Galaxy) at 5.9~Mpc, 
NGC~5457 (Pinwheel Galaxy, M101) at 6.7~Mpc, and NGC~5194 (Whirlpool Galaxy, M51) at 5.9~Mpc. 
These galaxies are the best sites to search for extragalactic GFs in addition to the 
previously mentioned galaxies: M82 at $d = 3.4$~Mpc, NGC~253 (Sculptor Galaxy) at 2.5~Mpc, 
NGC~4945 at 3.7~Mpc, and M83 at~3.7~Mpc~\citep{Popov2006}.

\section{THE SEARCH FOR SGR GIANT FLARES IN 
THE KONUS-\textit{WIND} SHORT GRB SAMPLE} \label{GF_search}

The \citet{Palshin2013} catalogue contains IPN localizations for 271 out of the 296 \kw short GRBs 
with a duration $T_{50}\le 0.6$~s~\citep{Svinkin_classification}. This sample contains 
23~GRBs classified as short GRBs with extended emission~(EE).
These bursts were observed by at least one other IPN spacecraft, enabling their 
localizations to be constrained by triangulation. We limit our search to 140 bursts 
with IPN error box area less than 10~deg$^{2}$ and search for overlaps between 
each IPN box and galaxies from our sample. In the search a galaxy 
is modelled, following~\citet{Ofek2007}, by a circle determined by the galaxy's major diameter from GWGC.

We found that 12 out of the 140 IPN error boxes, with a total area of 217~deg$^{2}$, overlap 
20 galaxies (none of the localizations of short GRBs with~EE contains a galaxy). 
This number is lower than that expected for chance coincidences calculated 
for the 140 IPN boxes (22--44 galaxies at 95\% CL). The confidence interval was 
calculated using the bootstrap approach, creating 1000 random realizations of the galaxy sample 
and counting the number of overlaps with the IPN boxes list for each realization. 
We used the 25th and 975th ranked values as the 95\% confidence interval boundaries. 

Only one of these error boxes (GRB20050312\_T20417\footnote{See the burst details in~\citet{Palshin2013}}) 
overlaps the outskirts of the Virgo cluster, but the box does not contain any 
galaxy from our sample. The cluster was modelled with a circle centred 
at $\rmn{R.A.}=188\degr$, $\rmn{Dec.}=12\degr$, whose radius 
is $6\degr$ using the cluster parameters from~\citet{Binggeli1987}.

We found that only two previously reported GF candidates, GRB~051103 associated 
with the M81/M82 group of galaxies (error box area $4.3\times10^{-3}$~deg$^{2}$) and GRB~070201 
associated with the Andromeda galaxy (error box area  0.123~deg$^{2}$), have low probabilities 
of chance coincidence ($P_{\rmn{chance}}\sim1$\%). $P_{\rmn{chance}}$ is the probability 
to find at least one galaxy in the given IPN error box. The probability was calculated in 
the same way as the confidence interval.

We then applied the search procedure to the sample of 98 bursts with IPN error box area 
less than 1~deg$^{2}$~\citep{Palshin2013} which contains short GRBs observed by 
at least one \textit{distant} spacecraft. We found that only the localizations of the two 
above-mentioned bursts contain galaxies.


A common feature of all known GFs is the short duration of the initial peak, $\la 500$~ms, and 
the short rise time, $t_{\rmn{r}} \la 25$~ms. Among 296 \kw short GRBs 40 have $t_{\rmn{r}}<25$~ms 
and a duration $<500$~ms. The previously discussed GF candidates GRB~051103 and GRB~070201 have 
$t_{\rmn{r}}=2$~ms and $t_{\rmn{r}}=24$~ms, respectively. We then limited our search to 17 of these 
bursts with IPN error box area less than 10~deg$^{2}$. 

We found that four out of the 17 IPN boxes, with a total area of 47~deg$^{2}$, contain five galaxies. 
This number is consistent with that expected for chance coincidences calculated for this sample: 
5--16 galaxies (95\% CL). Among these four bursts only GRB~051103 and GRB~070201 have 
low probabilities of chance coincidence. The results for all burst samples are 
summarized in Table~\ref{tab:SearchSummary}.

Given the product of the completeness factors $\epsilon_{\rmn{G}} \epsilon_{\rmn{L}} \epsilon_{\rmn{SN}} \approx 76$\%, 
and assuming that the search yielded two GF candidates among 98 well localized \kw 
short GRBs, the upper limit to the fraction of GF among \kw short GRBs is less than 8\% (=6.296/98/0.76)
at a one-sided 95\% CL~\citep{Gehrels1986}. Due to the whole sky coverage of the IPN this limit can be 
adopted for all short GRBs with fluences above $\sim 5\times 10^{-7}$~erg~cm$^{-2}$.
The resulting upper limit is stricter than the limit obtained in~\cite{Ofek2007}. 


\begin{table*}
\begin{minipage}{126mm}
\caption{Summary of GF search in the \kw short GRB sample}
\label{tab:SearchSummary}
\begin{tabular}{c c c c}
\hline
Sample        & Number   & Number of galaxies & Expected number of galaxies \\
description   & of boxes & in boxes           & in boxes at 95\% CL \\
\hline
IPN box area $<10$~deg$^2$  & 140 & 20 & 22--44 \\
IPN box area $<1$~deg$^2$   & 98 & 2 & 0--7\\
$t_{\rmn{r}} \leq 25$~ms, $T_{100}<500$~ms, & 17 & 5 & 5--16\\
and IPN box area $<10$~deg$^2$ & & & \\
\hline
\end{tabular}
\end{minipage}
\end{table*}

\section{THE LIMITS ON THE GIANT FLARE RATE} \label{GF_rate}
%

Assuming that only one GF with energy output $Q \ga 10^{46}$~erg has been 
observed, in the M81/M82 group of galaxies, within a volume $d \le 30$~Mpc, 
it is possible to calculate an upper limit to the rate of such GFs. 

We assume that the number of active SGRs ($N_{\rmn{SGR}}(d)$) in the galaxies within distance $d$  
is proportional to the core-collapse supernova rate $R_{\rmn{SN}}(d)=4/3 \pi d^3 r_{\rmn{SN}}$, 
where $r_{\rmn{SN}}$ is the volumetric CCSN rate.
\begin{equation}\label{eq:NumSGR}
N_{\rmn{SGR}} (d) = \frac{N_{\rmn{SGR}, \rmn{MW+LMC}}}{ R_{\rmn{SN}, \rmn{MW+LMC}}} R_{\rmn{SN}}(d).
\end{equation}
The Galactic CCSN rate is $R_{\rmn{SN}, \rmn{MW}} = 0.028\pm0.006$~yr$^{-1}$ with a 
systematic uncertainty of a factor of $\sim 2$~\citep{Li2011part3} and 
the LMC rate is $R_{\rmn{SN}, \rmn{LMC}} = 0.013\pm0.009$~yr$^{-1}$~\citep{Bergh1991}. 
Hence the total CCSN rate in MW and LMC is $R_{\rmn{SN}, \rmn{MW+LMC}} = 0.041\pm0.011$~yr$^{-1}$. 

The observed rate of GFs per SGR is given by a simplified version of eq.~3 from~\citet{Ofek2007}:
\begin{equation}\label{eq:RateGF}
R_{\rmn{GF}} = \frac{N_{\rmn{GF},\rmn{obs}}}{\Delta T N_{\rmn{SGR}}(d_{\rmn{max}})} ,
\end{equation}
where $N_{\rmn{GF},\rmn{obs}}$ is the number of observed GF, $\Delta T=16$~yr 
is the \kw observation time, and $N_{\rmn{SGR}}(d)$ is given by eq.~\ref{eq:NumSGR}. 
To estimate an upper limit to $R_{\rmn{GF}}$ in case of one observed GF 
we used the 95\% one-sided upper limit of $N_{\rmn{GF}, \rmn{obs}}=4.744$~\citep{Gehrels1986}.

For the rate of GFs with energy output $Q \ga 10^{46}$~erg eq.~\ref{eq:RateGF} 
gives an upper limit of
${(0.6\textrm{--}1.2)\times 10^{-4} Q_{46}^{-1.5}}$~yr$^{-1}$~SGR$^{-1}$, 
where $Q_{46}$ is a GF energy output in $10^{46}$~erg. 
The detection of only one GF with energy output $Q \ga 10^{46}$~erg in the last 35~yr 
(since 1979) from SGR~1806$-$20 implies that the Galactic rate is 
${(0.005\textrm{--}1)\times 10^{-2}}$~yr$^{-1}$~SGR$^{-1}$ ($=1_{-0.98}^{+4.6} / 35 / 15 $) 
at the one-sided 95\% CL. This value is consistent with the upper limit derived above for the 
SGR~1806$-$20 distance 9.4--18.6~kpc, primarily due to the large uncertainties 
in both the upper limit and the Galactic GF rate. 

For less energetic flares with energy output $Q \la 10^{45}$~erg which can be 
detected by \kw at distances up to 6.3~Mpc, assuming one such flare was 
detected from the Andromeda galaxy, the implied rate upper limit is 
${(0.9\textrm{--}1.7)\times 10^{-3}}$~yr$^{-1}$~SGR$^{-1}$ (95\% CL). 
This limit is consistent with the observed Galactic rate of 
${(0.05\textrm{--}1.4)\times 10^{-2}}$~yr$^{-1}$~SGR$^{-1}$ (95\%~CL).

\section{SUMMARY AND DISCUSSION} \label{Discussion}
We have estimated the \kw and IPN sensitivity to GFs and derived the limiting detection 
distance for GFs similar to that from SGR~1806$-$20 to be $(18\textrm{--}30) d_{15}$~Mpc. 
Less energetic flares such as those from SGR~1900+14 and SGR~0526$-$66 can be 
detected by \kw and IPN from galaxies at $d \la 6$~Mpc. 
We searched for \kw short GRBs that spatially coincide with galaxies 
within 30~Mpc and have found that only two previously reported GF candidates 
have a low chance coincidence probability: GRB~051103 and GRB~070201 
localized by the IPN in the M81/M82 group of galaxies and the Andromeda galaxy, respectively. 
We have not found any candidate GF from the Virgo cluster. 
Assuming only one GF with energy output $Q \ga 10^{46}$~erg was 
observed in the M81/M82 group of galaxies within the volume $d \le 30$~Mpc we obtain an 
upper limit to the rate of such flares to be 
${(0.6\textrm{--}1.2)\times 10^{-4} Q_{46}^{-1.5}}$~yr$^{-1}$~per~SGR, 
where $Q_{46}$ is the GF energy output in $10^{46}$~erg. 
This limit was calculated using the largest sample of well localized short GRBs 
and is consistent with the finding of~\citet{Ofek2007}.  The limit implies 
roughly one giant flare with energy output $Q \ga 10^{46}$~erg during 
the lifetime of an SGR, $10^3\textrm{--}10^5$~yr. 

For a GF with energy output of the order of the 5~March~1979 event ($Q \la 10^{45}$~erg) 
the implied rate is nearly an order of magnitude higher 
$(0.9\textrm{--}1.7)\times 10^{-3}$~yr$^{-1}$~SGR$^{-1}$ (95\% CL). This can be 
interpreted to mean that more than one such GF can be observed during SGR lifetime. 
The measured dipolar magnetic fields of SGR~1900+14 and SGR~0526$-$66, 
$5.6\times10^{14}$~G and $7\times10^{14}$~G, respectively~\citep{Olausen_Kaspi2014}, 
seem to be sufficient to power a dozen GFs with $Q \sim 10^{45}$~erg.

The upper limits we have presented contain uncertainties of about an order of 
magnitude which account for the uncertainty in the Galactic CCSN rate 
and the limiting detection distance of the IPN. 

We found galaxies which are promising targets for GF observations:  
PGC047885, IC~0342, NGC~6946, NGC~5457, and  NGC~5194, in addition to those
discussed in~\citet{Popov2006}.

We also derived the upper limit to the rate of bright GFs using \textit{Swift}-BAT data. 
Since its launch in November 2004 \textit{Swift} has observed only one candidate 
extragalactic GF, GRB~050906~\citep{Levan2008}, with a suggested host IC~328 
at $\approx 130$~Mpc. This burst had the lowest fluence in the 15--150~keV band 
of all the reported GRBs, $S_{\rmn{min}} = 6.1\times10^{-9}$~erg~cm$^{-2}$~\citep{Sakamoto2011ApJS}, 
and a soft spectrum described by a power law model with a photon index of~$-1.7$. 
The extrapolated energy output of the GF from SGR~1806$-$20 from 20~keV--10~MeV  
to 15--150~keV using a cutoff power law model with $\alpha=-0.73$ and $E_{\rmn{p}}=850$~keV 
yields $2.5\times10^{45}$~erg, which corresponds to a limiting distance of 60~Mpc. 
The interpretation of GRB~050906 as an SGR giant flare in IC 328 is unlikely 
due to the derived \textit{Swift}-BAT limiting detection distance.

The BAT 90\% sky exposure time during 2004--2010 was $7.25\times10^{6}$~s 
(0.23~yrs)~\citep{Baumgartner2013ApJS}, and extrapolation of this value to 2004--2013 
gives 0.35~yrs. The non-detection of a GF in the volume $d \la 60$~Mpc during 
\textit{Swift}-BAT operations implies a one-sided 95\% CL upper limit 
for a GF with this energy output to be $\sim 6 \times 10^{-4} Q_{45}^{-1.5} $~yr$^{-1}$~SGR$^{-1}$, 
where $Q_{45}$ is the GF energy output in $10^{45}$~erg in the 15--150~keV band. 
Despite BAT's high sensitivity this upper limit is less strict than that obtained using 
\kw and IPN data due to a lower period of whole sky observations. 
Nevertheless the \textit{Swift}-BAT is a very promising mission 
to detect extragalactic SGR giant flares.

\section*{ACKNOWLEDGEMENTS}
We are grateful to V.~D.~Pal'shin and R.~C.~Duncan for valuable discussions.
The Konus-Wind experiment is partially supported by a Russian Space Agency contract and 
RFBR grants 12-02-00032a and 13-02-12017-ofi-m. K.H. is grateful for IPN
support under the following NASA, JPL, and MIT grants and contracts. 
JPL 958056 and 1268385 (Ulysses); 
NNX07AH52G and NNX12AE41G (ADA and ADAP); 
NAG5-12614, NNG04GM50G, NNG06GE69G, NNX07AQ22G, 
NNX08AC90G, NNX08AX95G and NNX09AR28G (INTEGRAL); 
NNG05GTF72G, NNG06GI89G, NNX07AJ65G, NNX08AN23G, 
NNX09AO97G, NNX10AI23G, and NNX12AD68G (Swift); 
NAG5-3500 and NAG5-9503 (NEAR); 
MIT-SC-R-293291 and NAG5-11451 (HETE-2); 
JPL 1282043 (Odyssey); 
NNX06AI36G, NNX08AB84G, NNX08AZ85G, NNX09AV61G, NNX10AR12G (Suzaku); 
NNX09AU03G, NNX10AU34G, and NNX11AP96G (Fermi); 
NNX07AR71G (MESSENGER); 
NAG5-7766, NAG5-9126, and NAG5-10710 (BeppoSAX).

\label{lastpage}

\bibliographystyle{mn2e}
\bibliography{biblio_sgr_gf}

\end{document}